\documentclass[nohyper]{JHEP}
\usepackage{amsmath}
\usepackage{amssymb}
\usepackage{cite}

\usepackage{epsfig}

\psfull

\newcommand{\ie}{\emph{i.e.}\ }
\newcommand{\cnf}{\emph{cf.}\ }

\def\cO#1{{\cal{O}}\left(#1\right)}

\def\vkti{\vec{k}_{ti}}
\def\vkt{\vec{k}_{t}}
\def\vpt{\vec{p}_{t}}
\def\pt{{p}_{t}}
\def\vb{\vec{b}}

\def\kt{k_t}
\def\kti{k_{ti}}
\def\ktj{k_{tj}}

\def\da{\partial_a}

\def\hc{{\cal H}_{\mathrm{C}}}
\def\hr{{\cal H}_{\mathrm{R}}}
\def\rc{{\cal R}_{\mathrm{C}}}
\def\rr{{\cal R}_{\mathrm{R}}}

\def\elim{{\cal E}_{\mathrm{lim}}}

\def\vb{\vec{b}}

\def\SigmaTilde{{\widetilde \Sigma}}

\def\nubar{{\bar \nu}}
\def\bbar{{\bar b}}
\def\qbar{{\bar q}}

\newcommand{\bq}{\mathbf{q}}
\newcommand{\bP}{\mathbf{P}}
\newcommand{\bC}{\mathbf{C}}

\def\cR{{\cal{R}}}               
\newcommand{\dsig}{{d^2\!\sigma}}

\def\half{\mbox{\small $\frac{1}{2}$}}

\def\MSbar{\overline{\mbox{\scriptsize MS}}}
\def\DIS{\mathrm{DIS}}

\def\cf{C_F}
\def\tr{T_R}
\def\ca{C_A}
\def\nf{n_{\!f}}

\def\as{\alpha_{{\textsc{s}}}}
\def\gae{{\gamma_{\textsc{e}}}}
\def\asb{{\bar \alpha}_{{\textsc{s}}}}

\def\ee{e^+e^-}

\newcommand\epjcd[3]  {
                {{\it Eur.\ Phys.\ J. Direct }{\bf C #1} (#2) #3}}

\title{Resummation of thrust distributions in DIS\thanks{Research
    supported in part by MURST, Italy and by the EU Fourth Framework
    Programme `Training and Mobility of Researchers', Network `Quantum
    Chromodynamics and the Deep Structure of Elementary Particles',
    contract FMRX-CT98-0194 (DG 12-MIHT).}}

\author{Vito Antonelli, Mrinal Dasgupta \\
  Dipartimento di Fisica, Universit\`a di Milano Bicocca
  and INFN, Sezione di Milano, Italy}
\author{Gavin P. Salam \\
CERN, TH Division, 1211 Geneva 23, Switzerland}

\abstract{We calculate the resummed distributions for the thrust in
  DIS in the limit $T\to1$. Two variants of the thrust are considered:
  that normalised to $Q/2$, and that normalised to the energy in the
  current hemisphere. The results expanded to second order are
  compared to predictions from the Monte Carlo programs DISENT and
  DISASTER++.  A prescription is given for matching the resummed
  expressions with the full fixed order calculation.}

\keywords{QCD, NLO Computations, Jets, Deep Inelastic Scattering}

\preprint{Bicocca--FT--99--32 \\
  CERN--TH/99--396\\
  hep-ph/9912488 \\
  December 1999
  }

\begin{document}

\section{Introduction}

For some time now it has been standard practice in $e^+e^-$ reactions
to compare event-shape distributions with resummed perturbative
predictions (see for instance \cite{CTTW}). The resummation is
necessary because in the two-jet limit (small values of the shape
variable) the presence of large logarithms spoils the convergence of
the fixed-order calculations.  Such resummed analyses have led to
valuable information about the strong coupling constant and also about
non-perturbative effects \cite{eeExp}.

At HERA, similar studies of DIS event-shape distributions are being
carried out by both collaborations \cite{H1dists,ZEUSdists}, but as
yet no perturbative resummed calculations exist for comparison between
data and theory.  Here we present the results of such a calculation.

So far resummed predictions for processes involving initial state
hadrons exist for observables such as DIS and Drell-Yan cross-sections
in the semi-inclusive limits where $x$ for DIS and $\tau$ for
Drell-Yan are close to one \cite{Sterman,CatTrent,CMW}. In addition
there are predictions for jet-multiplicities in DIS \cite{CDW} and the
W transverse momentum in Drell-Yan production \cite{Wkt} which
entail somewhat similar considerations to those that are encountered
here.

In DIS, event shapes are defined in the current hemisphere of the
Breit frame to reduce contamination from remnant fragmentation, which
is beyond perturbative control \cite{BF}. The distribution of partons
in the current hemisphere is analogous to that in a single hemisphere
in $e^+e^-$, enabling us to adopt the usual $e^+e^-$ methods for the
resummation. But it turns out that event-shape variables also depend
on emissions in the remnant hemisphere (typically through recoil
effects) --- a proper resummed treatment of the space-like branching
of the incoming parton is therefore required.

The fact that only one, predefined, hemisphere is used in DIS event
shapes leads to a number of subtleties compared to $\ee$ event shapes.
Event shapes measure properties of the energy-momentum flow in the
current hemisphere. They are always dimensionless, so it is necessary
to normalise this measure of energy-momentum flow. Two choices are
possible: a normalisation to $Q/2$, or to $E$, the energy in the
current hemisphere (whereas in $\ee$ the total energy $E$ is equal to
$Q$). Furthermore, one often requires the choice of an axis along
which to project the momentum flow (for example for the thrust, the
broadening). In $\ee$ there is one logical choice of axis, namely the
thrust axis (that which maximises the thrust). In DIS, two axes arise
naturally: the photon axis and the thrust axis (which is given by the
direction of the sum of all momenta in the current hemisphere).

In this article we shall consider two variants of the thrust. Both are
measured with respect to the photon axis: one, $T_Q$, is normalised to
$Q/2$ and the other, $T_E$, is normalised to $E$. In the limit of a
1+1 jet event both tend to $T=1$.  Since we are particularly
interested in this limit, it will be convenient to define $\tau=1-T$
and consider the limit of small $\tau$.  While we are generally
interested in the differential distribution $\frac{1}{\sigma}\frac{d
  \sigma}{d\tau}$, it will turn out to be more convenient to look at
the integrated event-shape cross section
\begin{equation}
\Sigma(\tau) = \int_{0}^{\tau}d\tau^{\prime}
\frac{d\sigma}{d\tau^{\prime}},
\end{equation}
from which the distribution can be straightforwardly obtained by
differentiation.

For small $\tau$ one finds that the shape cross-section contains terms
$\as^n \ln^m \tau,\; m \leq 2n$ coming from multiple soft and
collinear radiation. The $\ln \tau$ factors compensate the smallness
of $\as$, hence the need arises for resummation, which leads to a
result schematically of the form $\Sigma(\tau) \sim \exp(- G_{12}
\as \ln^2 \tau + \ldots)$, with $G_{12}$ some number which will be
calculated. From the point of view of classifying the terms that we
shall calculate, it turns out to be most convenient to consider $\ln
\Sigma$. This contains terms $\as^n \ln^m \tau$ with $m\le n+1$. Terms
with $m=n+1$ are referred to as leading logarithmic (LL), while those
with $m=n$ are known as next-to-leading logarithmic (NLL). We shall
resum both these sets of terms, and neglect as subleading the
remaining terms, $m<n$.

A word of caution is needed about $\tau_E$, and in general about all
event-shape measures which are normalised to the energy in the current
hemisphere, $E$. At $\cO{\as}$ there are configurations in which the
current hemisphere is empty. At higher orders it can be filled by
soft-gluon radiation from the partons in the remnant hemisphere. If
the event-shape is normalised to $E$ then it can take on finite
(significantly different from 0 and/or 1) values due to the presence
of soft-radiation; in other words it will be infrared unsafe starting
from $\cO{\as^2}$. This has been known for some time. The standard
experimental practice is to exclude events in which the energy in the
current hemisphere is too low.  For example, H1 set the threshold
$E>\elim= Q/10$ \cite{H1dists}.  Though formally this eliminates any
infrared unsafety, one is still left with potentially large logs %
$\as (\as \ln Q/2\elim)^n$.  Thus we recommend that $\elim$ be chosen
somewhat larger, say $\elim=Q/4$. Though this will slightly reduce the
statistics, it ought to improve the convergence of the perturbation
series.

The structure of this article is as follows. In section
\ref{sec:kindef} we discuss the kinematics and define the thrust
measures which will be studied. In section~\ref{sec:resum} we perform
the resummations. Section~\ref{sec:finalres} gives the expressions to
be used for the practical calculation of resummed distributions. The
results are expanded to second order and compared to the predictions
from the fixed order Monte Carlo programs DISASTER++ \cite{DR} and
DISENT \cite{DT}. This enables us to cast some light on the
differences between DISASTER++ and DISENT, observed in \cite{mccance}.
Finally a \textit{matching} prescription is proposed for combining the
resummed and fixed order predictions. This is followed by the
conclusions.

The appendices provide details about the leading order calculation of
the thrust (appendix~\ref{app:calcLO}), the techniques used for the
inverse Mellin transforms (appendix~\ref{app:invmellin}) and the
integrations needed for a result correct to NLL
(appendix~\ref{app:twoloopR}).

\section{Thrust definitions and kinematics}
\label{sec:kindef}

\FIGURE{\epsfig{file=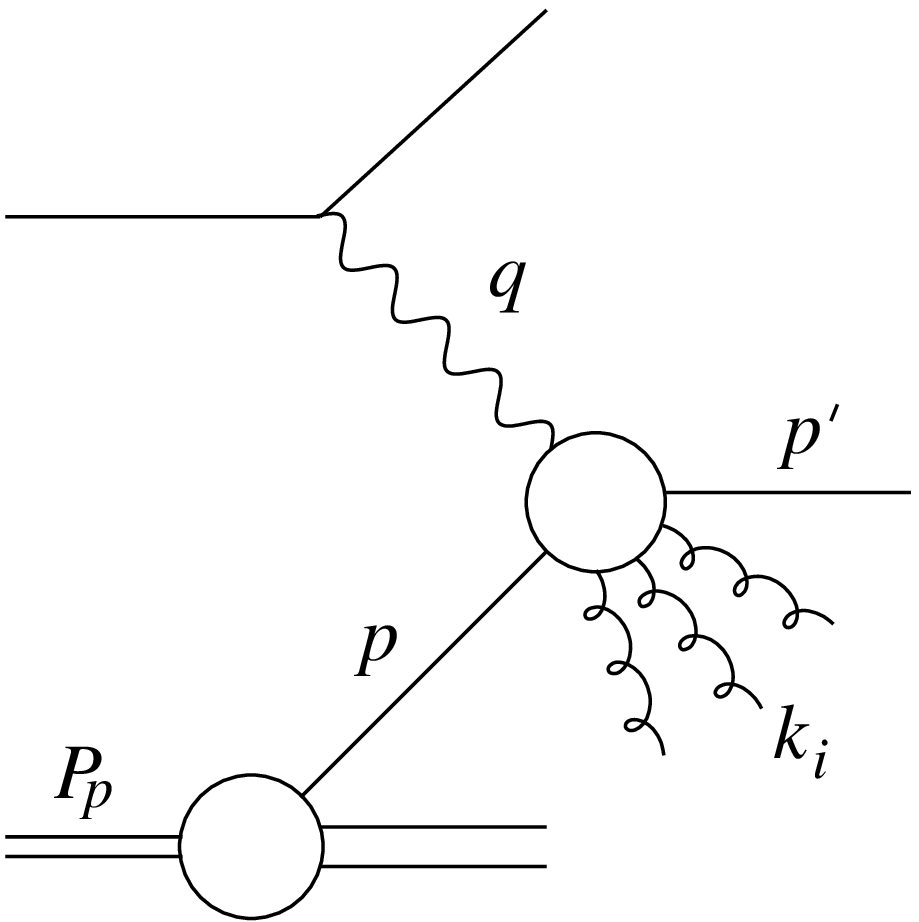,width=0.25\textwidth}\\
  \vspace{0.05cm}
  \label{fig:kin}
  \caption{Kinematics.}}
It is convenient to write the momenta $k_i$ of radiated partons
(gluons and/or quark-antiquark pairs) in terms of Sudakov
(light-cone) variables as (Fig.~1)
\begin{equation}\label{eq:suddef}
 k_i\>=\>  \alpha_i P \>+\>\beta_i P' \>+\>k_{ti}\>,
\qquad\qquad \alpha_i\beta_i\>=\>\vec k_{ti}^2/Q^2
\end{equation}
where $P$ and $P'$ are light-like vectors
along the incoming parton and current directions,
respectively, in the Breit frame of reference:
\begin{equation}
P\>=\>x P_p\>,\>\>
P'\>=\>x P_p+q\>,\>\> 
2(P\cdot P')\>=\>-q^2\>\equiv\>Q^2
\end{equation}
where $P_p$ is the incoming proton momentum and
$x=Q^2/2(P_p\cdot q)$ is the Bjorken variable. 
Thus in the Breit frame we can write
$P=\half Q(1,0,0,-1)$ and $P'=\half Q(1,0,0,1)$,
taking the current direction as the $z$-axis. Particles
in the {\em current hemisphere} $\hc$ have $\beta_i>\alpha_i$
while those in the proton {\em remnant hemisphere} $\hr$ have
$\alpha_i>\beta_i$.

The momenta $p$ and $p'$ of the initial- and final-state quarks
can also be resolved along the Sudakov vectors  $P$ and $P'$.
From momentum conservation,

\begin{align}
  p  &= \left(1+\alpha'+\sum\alpha_i\right)P\,,\nonumber \\
  p' &= \alpha' P+\left(1-\sum\beta_i\right)P'-\sum k_{ti}\,,
\end{align}
where
\begin{equation}\label{eq:alprime}
\alpha'\left(1-\sum\beta_i\right)\>=\>\left(\sum\vec
  k_{ti}\right)^2/Q^2\;. 
\end{equation}
The initial quark is assumed to be collinear with the proton
direction and is therefore aligned along $P$ (neglecting possible
`intrinsic' transverse momentum).

\subsection{Thrust normalised to $Q/2$, $T_Q$}
Consider first the thrust $T_Q$, defined by the sum of longitudinal
momenta in the current hemisphere normalised to $Q/2$:
\begin{equation}
T_Q =\frac 2 Q \sum_{a\in \hc}p_{za}\>=\>
\frac 2 Q \left(p'_z+\sum_{i\in\hc} k_{zi}\right),
\end{equation}
where the index $a$ run over all partons and the the index $i$
runs only over emitted partons.  We have assumed that the outgoing
quark momentum $p'$ lies in the current hemisphere, because the
consideration of the non-soft, non-collinear emissions that would
result in this not being true is beyond our concern in the present
case. We have
\begin{align}
\label{eq:tauQ}
\tau_Q\equiv 1-T_Q &= \sum\beta_i
-\sum_{\hc}(\beta_i-\alpha_i)+\alpha'
\nonumber \\
&= \sum_{\hr}\beta_i +\sum_{\hc}\alpha_i+\alpha'
\>=\>\sum\min\{\alpha_i,\beta_i\}+\alpha'\;.
\end{align}
Where not explicitly stated, the sums extend over all emitted partons,
irrespective of which hemisphere they are in. Here the contribution
from those partons that are not directly in the current hemisphere
arises through the recoil effect that they have on the current quark.

\subsection{Thrust normalised to $E$, $T_E$}

Another way to define the current jet thrust is to normalise it to the
total energy in the current hemisphere, instead of $Q/2$. This gives
\begin{equation}
T_E = T_Q/(1-\varepsilon)\,,
\end{equation}
where 
\begin{equation}\label{eq:epsdef}
\varepsilon  \equiv 1-\frac 2 Q \sum_{a\in \hc}E_a\;.
\end{equation}
The quantity $\varepsilon$, called the energy deficit in the current
hemisphere, is itself an interesting shape variable \cite{DWmilan}.
It is given by
\begin{equation}\label{eq:epsbet}
\varepsilon
= \sum\beta_i -\sum_{\hc}(\beta_i+\alpha_i)-\alpha'
\>=\>\sum_{\hr}\beta_i -\sum_{\hc}\alpha_i-\alpha'\;.
\end{equation}
Hence, we have,
\begin{equation}
  \label{eq:tauE}
  \tau_E = 1 - T_E = \frac{2}{1 -
    \sum_{\hr}\beta_i +\sum_{\hc}\alpha_i+\alpha'} \left(
    \sum_{\hc} \alpha_i + \alpha'\right)\,.
\end{equation}

\section{Resummation}\label{sec:resum}

We shall consider a reduced integrated cross section, where by reduced
we mean that it will be the contribution to $F_2$ (rather than to
$d\sigma /dxdQ^2$) which comes from configurations with $1-T<\tau$.
There will be no need to consider contributions to $F_L$ since they
are suppressed by powers of $\tau$ for small $\tau$.

We write the following expression for the (reduced) cross section for
the emission of $m$ gluons in the remnant hemisphere and $n$ in the
current hemisphere, where all the gluons have $k_t \ll Q$,
\begin{equation}
  \label{eq:right}
  e_q^2 q_N(Q_0^2) \cdot \frac{1}{m!} \prod_{i}^{m} 
  \int_{Q_0^2}^{Q^2} \frac{d \kti^2}{\kti^2} \frac{\as(\kti^2) \cf}{2\pi}
   \rho_{R,i} \,\cdot\,
  \frac{1}{n!} \prod_{j}^{n} 
  \int^{Q^2} \frac{d \ktj^2}{\ktj^2} \frac{\as(\ktj^2) \cf}{2\pi}
   \rho_{C,j}\,.
\end{equation}
with
\begin{subequations}
  \label{eq:RrRt}
\begin{align}
  \label{eq:Rr}
  \rho_{R,i} &= \int z_i^N \; dz_i \,\frac{1 + z_i^2}{1-z_i}\,
  \Theta(Q\alpha_i- k_{ti}), \\
  \rho_{C,j} & = \int \frac{d \beta_j}{\beta_j} \left (1+(1-\beta_j)^2
  \right ) \Theta(Q\beta_j-\ktj).
\end{align}
\end{subequations}
$N$ is the moment variable (conjugate to Bjorken-$x$) and
\begin{equation}
\alpha_i = \frac{1-z_i}{z_1 \ldots z_i}\,. \label{eq:alphai}
\end{equation}
We have introduced a factorisation scale $Q_0^2$, and
accordingly in $\hr$ consider only emissions above that scale. $Q_0^2$
is chosen so as to be much smaller than any other scale in the
problem. 

For the event shapes considered here, the independent gluon emission
pattern written above is enough to reproduce the behaviour of the full
QCD matrix elements up to the required (NLL) accuracy, with the
following restrictions: $\as$ must be evaluated to two loops in the
CMW (or Physical) scheme \cite{CMW,DKT}; we must also put in the virtual
corrections and take into account all possible branchings of the
incoming leg, not just the emission of gluons from a quark; but these
can both be done trivially later on. It is not necessary to take into
account the splitting of emitted partons, other than what is already
implicitly included through the running of the coupling.

The fundamental difference between the expression given above and the
corresponding one for $\ee$ comes from the presence of the quark
distribution. Hard collinear emissions change the momentum fraction of
the incoming parton, and this must be accounted for, through a
(multiple) convolution of some function (to be determined) with the
parton distributions. We take moments ($N$) with respect to $x$ in
order convert that convolution into a product. These moments enter
only for the emissions in $\hr$ --- emissions in $\hc$ are essentially
identical to those in a hemisphere of $\ee$.

The use of moments does not however remove all difficulties. In
particular in eq.~\eqref{eq:alphai} we are still left with a product
of $z_i$'s. This matters both in the $\Theta$ function in
eq.~\eqref{eq:Rr} and in the contribution to the thrust, which for say
$\tau_Q$ will be $\beta_i = k_{ti}^2/(Q^2\alpha_i)$ (\cnf 
eq.~\eqref{eq:tauQ}).  The solution relies on the fact that the gluons
which contribute to the thrust are either the gluon with the highest
transverse momentum or soft gluons.  Considering the gluons as ordered
in transverse momentum, the gluon with highest transverse momentum is
number 1 and $\alpha_1=(1-z_1)/z_1$. One can equally well consider the
gluons to be ordered in angle, and if gluon $m$ has the highest
transverse momentum, then gluons with $i<m$ must have $1- z_i \ll 1$
and one can write $\alpha_m \simeq (1-z_m)/z_m$. For soft gluons we
can approximate the splitting function by $2/(1-z_i)$, and then
replace $(1-z_i)/(z_1 \ldots z_{i-1})\Rightarrow(1-\zeta_i)$. The
integration measure remains $2d\zeta_i/(1-\zeta_i)$, and $\alpha_i$
just becomes $(1-\zeta_i)/\zeta_i$. So we can in general just replace
$\alpha_i$ by $(1-z_i)/z_i$ (where $z_i$ should sometimes be understood to
mean $\zeta_i$).

\subsection{Resummation of  $\tau_Q$}
\label{sec:tauQ}

To write a resummed expression we take the Mellin transform of our
$\Theta$ function for $\tau_Q$:
\begin{equation}
  \label{eq:step}
\Theta\left(\tau - \sum_{\hr}\beta_j
  -\sum_{\hc}\alpha_i-\alpha'\right) = 
\int \frac{d\nu}{2\pi i \nu} e^{\tau \nu} 
\left( \prod_{\hr} e^{-\nu \beta_i} \right)
\left( \prod_{\hc} e^{-\nu \alpha_i} \right)
e^{-\nu \alpha'}.
\end{equation}
This almost factorises the expression for the $\Theta$ function into
pieces which each depend on a single emission. There remains the less
trivial term $e^{-\nu\alpha'}$: from \eqref{eq:alprime} we know that
$\alpha'$ is approximately (using $1-\sum \beta\simeq1$) the
squared vector sum of emitted transverse momenta. For now it is
actually sufficient to consider the sum of squared momenta. For soft
and collinear gluons $k_{ti}^2/Q^2 = \alpha_i \beta_i $ is negligible
compared to $\min(\alpha_i,\beta_i)$. Only for collinear but hard
gluons is $k_{ti}^2/Q^2$ comparable to $\min(\alpha_i,\beta_i)$
because one of the $\alpha_i, \beta_i$ is $\cO{1}$. In the current
hemisphere, since $\beta_i <1$, the contribution from the $k_t^2/Q^2$
can be neglected: the region where it is comparable to $\alpha_i$ is
single-logarithmic and its inclusion modifies $\tau$ by a numerical
factor, \ie changes $\as \ln \tau$ to say $\as \ln 2 \tau$, a
difference which is NNLL and so negligible. The difference is
similarly NNLL in the remnant hemisphere, however the fact that
$\alpha_i$ can be much greater than $1$ (as large as $1/x$), and hence
$k_{ti}^2/Q^2 \gg \beta_i$, means that one should at least keep track
of it.  Therefore in the remnant hemisphere contribution we retain a
contribution to $\tau$ of the form $k_t^2/Q^2$ (keeping track of it
more accurately would involve using the squared vector sum --- but for
$\tau_Q$ the error is down by one more logarithm). Thus we replace
\begin{equation}
\left( \prod_{\hr} e^{-\nu \beta_i} \right)
\left( \prod_{\hc} e^{-\nu \alpha_i} \right)
e^{-\nu \alpha'}
\Rightarrow 
\left( \prod_{\hr} e^{-\nu (\beta_i + k_{ti}^2/Q^2)} \right)
\left( \prod_{\hc} e^{-\nu \alpha_i} \right)\,.
\end{equation}

We then sum over $m$ and $n$ in \eqref{eq:right} to give us the
following exponentiated form for the integrated thrust distribution
\begin{equation}
  \label{eq:sigma}
  \Sigma_N(\tau) = e_q^2 q_N(Q^2) \int \frac{d\nu}{2\pi i \nu} \,
  e^{\tau \nu}\, 
  e^{-\rr(\nu) - \rc(\nu)}\,.
\end{equation}
The current hemisphere radiator, $\rc$ is 
\begin{equation}
  \label{eq:currentbase}
  \rc(\nu) = -
  \int \frac{d \kt^2}{\kt^2} \frac{\as(\kt^2)
      \cf}{2\pi} \int_0^1 d\beta\, \frac{1 + (1-\beta)^2}{\beta}\,
    \Theta(Q\beta - k_t) 
     \left( e^{-\nu \alpha} - 1 \right),
\end{equation}
where the $-1$ accounts for virtual corrections (as in
\cite{DLMSbroad}). The remnant hemisphere radiator is a little
trickier. The attentive reader will have noticed that
eq.~\eqref{eq:right} has a leading factor of $q_N(Q_0^2)$, whereas
eq.~\eqref{eq:sigma} has $q_N(Q^2)$.  Thus $\rr(\nu)$ should contain a
($\nu$-independent) piece to take into account the change in scale of
the quark distribution,
\begin{equation}
  \label{eq:changescale}
  \ln \frac{q_N(Q^2)}{q_N(Q_0^2)} = \int_{Q_0^2} ^{Q^2} \frac{d
      \kt^2}{\kt^2} \frac{\as(\kt^2) 
      }{2\pi} \gamma_{qq}(N),
\end{equation}
(where $\gamma_{qq}$ is the standard quark anomalous dimension) as
well as the part arising from the sum over $m$ in
eq.~\eqref{eq:right}. For now we are still working in a framework
involving only gluon emission from a quark, hence the use of only the
$\gamma_{qq}$ part of the anomalous dimension matrix to change the scale
of the quark distribution.

Therefore we have 
\begin{multline}
  \rr(\nu) = -\int_{Q_0^2}^{Q^2} \frac{d
      \kt^2}{\kt^2} \frac{\as(\kt^2) 
      \cf}{2\pi} \bigg( 
      \int_0^1 dz\,  \frac{1+z^2}{1-z} \Theta(Q\alpha - k_t)
      \left( z^N e^{-\nu (\beta + k_t^2/Q^2)} - 1 \right)\\
      - \frac{\gamma_{qq}(N)}{\cf}\bigg),
\end{multline}
where as before we have introduced a term $-1$ to account for virtual
corrections.  Next, we replace $\alpha=(1-z)/z$. In the $\Theta$
function we can neglect the $z$ in the denominator (the $\Theta$
function is relevant only for small $1-z$). So we obtain
\begin{multline}
  \rr(\nu) = -\int_{Q_0^2}^{Q^2} \frac{d
      \kt^2}{\kt^2} \frac{\as(\kt^2) 
      \cf}{2\pi} \bigg( 
      \int_0^1 dz\,  \frac{1+z^2}{1-z} \Theta(1-z - k_t/Q)
      \left( z^N e^{-\frac{\nu  k_t^2}{(1-z)Q^2}} - 1 \right)\\
      - \frac{\gamma_{qq}(N)}{\cf}\bigg).
\end{multline}
We note that in the limit of small $z$ the combination $(\beta +
\kt^2/Q^2)$ just reduces to $\kt^2/Q^2$, \ie a limit on $\tau$ is
just equivalent to a limit on the emitted transverse momentum.  Then
we write
\begin{equation}
  \label{eq:gaqq}
  \gamma_{qq}(N) = \cf\int_0^1 dz \frac{1+z^2}{1-z} (z^N  - 1) \Theta(1-z -
  k_t/Q) + \cO{\frac{k_t}{Q}},
\end{equation}
to obtain
\begin{equation}
  \label{eq:rrTauQmdl}
  \rr(\nu) = -\int^{Q^2} \frac{d
      \kt^2}{\kt^2} \frac{\as(\kt^2) 
      \cf}{2\pi} 
      \int_0^1 dz\,  z^N \frac{1+z^2}{1-z} \Theta(1-z - k_t/Q)
      \left( e^{-\frac{\nu  k_t^2}{(1-z)Q^2}} - 1 \right).
\end{equation}
The result no longer depends on our original choice of factorisation
scale, $Q_0^2$, and thus we drop the lower limit on the $k_t$
integral.  Our next step is to make the approximation, valid to our
accuracy \cite{DLMSbroad},
\begin{equation}
\label{eq:aprxthet}
  \left( e^{-\frac{\nu k_t^2}{Q^2(1-z)}}- 1 \right) \simeq
  -\Theta \left(\nubar - \frac{Q^2(1-z)}{\kt^2} \right)\,
\end{equation}
with $\nubar = \nu e^{\gae}$.  The $k_t$ integral can then be divided
into two pieces (according to whether the limit on $1-z$ in the
$\Theta$-function of \eqref{eq:aprxthet} is above or below $1$)
\begin{multline}
  \label{eq:remnantalmostthere}
  \rr(\nu) = 
   \int_{Q^2\nubar^{-1}}^{Q^2} \frac{d \kt^2}{\kt^2} \frac{\as(\kt^2)
      \cf}{2\pi} \int_0^{1-\kt/Q} dz\, z^N \frac{1+z^2}{1-z} \\
 + \int^{Q^2\nubar^{-1}}_{Q^2\nubar^{-2}} \frac{d \kt^2}{\kt^2}
   \frac{\as(\kt^2) 
      \cf}{2\pi} \int_{1-\nubar \kt^2/Q^2}^{1-\kt/Q} dz\,\frac{2}{1-z} 
 \,,
\end{multline}
where in the second term we have made the approximation $1-z\ll 1$. We 
express the $z$ integral in the first term as 
\begin{equation}
  \int_0^1 dz \frac{1+z^2}{1-z} (z^N - 1) + \int_0^{1-\kt/Q} dz
  \frac{1+z^2}{1-z} = \frac{\gamma_{qq}(N)}{C_F} + 2\ln \frac{Q}{k_t} - \frac32\,,
\end{equation}
where we have neglected terms of order $k_t/Q$. The $z$ integration in
the second term of \eqref{eq:remnantalmostthere} gives $2\ln(\nubar\kt/Q)$.

The current hemisphere radiator can be evaluated in a similar manner,
but without the complications from $N$ dependence. The sum of the
radiators is
\begin{multline}
  \label{eq:Rsum}
  \rr + \rc = \int_{Q^2\nubar^{-1}}^{Q^2} \frac{d \kt^2}{\kt^2}
  \frac{\as(\kt^2) \cf}{2\pi} \left(\frac{\gamma_{qq}(N)}{\cf} + 4\ln
    \frac{Q}{k_t}
    - 3 \right) \\
  +\int^{Q^2\nubar^{-1}}_{Q^2\nubar^{-2}} \frac{d \kt^2}{\kt^2}
  \frac{\as(\kt^2) \cf}{2\pi}\, 4\ln (\nubar\kt/Q) \,.
\end{multline}
We see that the quark anomalous dimension $\gamma_{qq}(N)$ appears in
our answer. It arises from the restriction of transverse momenta of
hard emissions. A consideration of all branchings leads to
$\gamma_{qq}(N)$ being replaced by the full anomalous dimension
matrix. This is a standard result in a situation where a restriction
is imposed on the final-state transverse momentum \cite{Wkt}. The
inclusion of all possible splittings leads one, however, to consider
also soft gluons emitted from a gluon.  Fortunately they can be
ignored in the present situation because it can be shown that for such
a gluon to contribute to the thrust it would have to be emitted at a
larger angle than the last hard emission, and thus, by coherence
\cite{coherence} it is emitted with the colour charge of a quark. To
see that the angle really is larger, let us consider two gluons
contributing equally to the thrust (in the remnant hemisphere), \ie 
with similar $\beta$ values. The angle of the gluon with respect to
the incoming quark is proportional to $\beta/\kt$. Gluons with
transverse momenta smaller than that of the last hard emission and
contributing equally to the thrust are bound therefore to have a
larger angle.

Thus our final answer in $\nu$ space is 
\begin{multline}
  \SigmaTilde_N(\nu) \simeq e_q^2 q_N(Q^2/\nubar)
  \exp\left[-
    \int_{Q^2\nubar^{-1}}^{Q^2} \frac{d \kt^2}{\kt^2} \frac{\as(\kt^2)
      \cf}{2\pi} \left(4\ln \frac{Q}{k_t} - 3 \right) \right. \\-
\left.  \int^{Q^2\nubar^{-1}}_{Q^2\nubar^{-2}} \frac{d \kt^2}{\kt^2}
  \frac{\as(\kt^2) 
      \cf}{2\pi}\,  4\ln(\nubar\kt/Q)
\right],
\end{multline}
where the integral involving the anomalous dimension matrix has been
absorbed into a rescaling of the quark structure function. For brevity
we rewrite the contents of the square bracket as $-2 R(\nubar)$.

The inverse Mellin transform is performed using the techniques of
\cite{DMS}, as outlined in appendix~\ref{app:invmellin} and gives
\begin{align}\label{tauqres}
  \Sigma_N(\tau_Q) &\simeq e_q^2 q_N(\tau_Q Q^2) \frac{1}{\Gamma(1 + 2R')} 
  \exp\left[-2R\left(\frac{e^{\gae}}{\tau_Q} \right)
  \right]\\
  \label{eq:tauqFinal}
  &\simeq
  e_q^2 q_N(\tau_Q Q^2) \frac{1}{\Gamma(1 + 2R')} 
  \exp\left[-2R\left(\frac{1}{\tau_Q}\right)  - 2\gae R' 
  \right],
\end{align}
where $R'=R'(1/\tau_Q)$. In the scale of the quark distribution,
numerical factors such as $e^\gae$ have been thrown away since they
correspond to NNLL effects.  Explicit expressions for $R$ and $R'$ are
given in appendix~\ref{app:twoloopR}.


\subsection{Resummation of $\tau_E$}

Starting from the expression for $\tau_E$, \eqref{eq:tauE}, we have
the $\Theta$-function
\begin{equation}
  \Theta\left( \frac{\tau}{2} - \sum_{\hc} \alpha_i -\left|\sum_{\hr}
      \vkti\right|^2 /Q^2\right).
\end{equation}
We have taken the contribution to $\alpha'$ only from transverse
momenta in the remnant hemisphere, because all transverse momenta in
the current hemisphere contribute much less (except for emissions with
$\beta\simeq 1$ and $\alpha \simeq \tau$, but this region is not
enhanced by any logs and so can be neglected).

We introduce 
\begin{equation}
  \vpt = \sum_{\hr} \vkti\,,
\end{equation}
and write the $\Theta$ function as
\begin{equation}
\int \frac{d\nu}{2\pi i \nu} e^{\tau \nu/2} 
\left( \prod_{\hc} e^{-\nu \alpha_i} \right)
\int\frac{d^2b\, d^2\pt}{(2\pi)^2}
e^{i\vb.\vpt}
\left( \prod_{\hr} e^{-i\vb. \vkti} \right)
e^{-\nu \pt^2/Q^2}.
\end{equation}
Analogously to what was done in the previous subsection, we write the
integrated cross section as
\begin{equation}
  \label{eq:SigmaTauE}
  \Sigma_N(\tau_E) = e_q^2 q_N(Q^2)\int \frac{d\nu}{2\pi i \nu}
  e^{\tau \nu/2} 
 \int \frac{d^2\pt\, d^2b}{(2\pi)^2} e^{-\nu
  \pt^2/Q^2} e^{i\vb.\vpt}    e^{-\rr(b) - \rc(\nu)}\,.
\end{equation}
Since the contribution from partons in the current hemisphere is the
same as in the $\tau_Q$ case (except for the factor of 2 included in
the definition of the Mellin transform), $\rc$ is the same as in that
case, \ie $\rc(\nu) = R(\nubar)$.

Arguments analogous to those used for $\tau_Q$,
eqs.~\eqref{eq:changescale} to \eqref{eq:rrTauQmdl}, lead us to the
following result for the remnant-hemisphere radiator $\rr(b)$,
\begin{align}
  \rr(b) &\simeq -\frac{\cf}{2\pi} \int^{Q^2} \frac{d^2\vec\kt}{\pi\kt^2}
  \,\as(\kt) \,  
  \int dz\, z^N    
  \frac{1+z^2}{1-z}\Theta(1-z - k_t/Q)\left( e^{-i \vb. \vkt} - 1\right)\\
   &\simeq \frac{\cf}{2\pi
     }\int^{Q^2}\frac{d\kt^2}{\kt^2} \,\as(\kt) \,\left(
     \frac{\gamma_{qq}(N)}{\cf}  + 2\ln
     \frac{Q}{k_t} - \frac{3}{2}\right)\left( 1 - J_0(b k_t)\right).
\end{align}
From \cite{DLMSbroad} this integral can be evaluated to the required
accuracy by replacing
\begin{equation}
  \left( 1 - J_0(b \kt)\right) \Rightarrow \Theta(\bbar \kt  - 1), \qquad
  \bbar \equiv \frac{e^\gae b}{2}.
\end{equation}
Thus we have 
\begin{equation}
  \rr(b) = R_U(\bbar^2 Q^2) + \int_{1/\bbar^2}^{Q^2}
  \frac{d\kt^2}{\kt^2} \frac{\as(k_t^2)}{2\pi}\gamma_{qq}(N),
\end{equation}
where $R_U(\nu)$ is defined in appendix~\ref{app:twoloopR}. We now
perform the $p$-integration in \eqref{eq:SigmaTauE} to get
\begin{align}
   \SigmaTilde_N(\nu) &\simeq  e_q^2 q_N(Q^2) \int \frac{d^2b}{2\pi}
   \frac{1}{2\nu} 
   e^{-b^2Q^2/4\nu}  \exp\left[-R_U(\bbar^2 Q^2) - \int_{1/\bbar^2}^{Q^2}
  \frac{d\kt^2}{\kt^2}\gamma_{qq}(N) - R(\nubar)\right]\\ &\simeq
    e_q^2 q_N\left(\frac{1}{\bbar_0^2}\right) \int \frac{d^2b}{2\pi}
    \frac{1}{2\nu}     e^{-b^2Q^2/4\nu}  
    \left(\frac{b^2}{b^2_0}\right)^{-R'_U}
      e^{-R_U(\bbar_0^2) - R(\nubar)} ,
\end{align}
where we have expanded $R_U(\bbar^2) = R_U(\bbar_0^2) + \ln b^2/b_0^2
\,R'_U(\bbar_0^2) + \cO{\as}$ and substituted $\gamma_{qq}$ with the full
anomalous dimension matrix and used it to change the scale of the
quark distribution. The integration over $b$ yields
\begin{equation}
  \SigmaTilde_N(\nu) \simeq e_q^2 q_N\left(\frac{1}{\bbar_0^2}\right)
  \left(\frac{b_0^2Q^2}{4\nu}\right)^{R'_U} 
  \Gamma\left(1 - 
    R_U'\right) e^{-R_U(\bbar_0^2 Q^2) - R(\nubar)} ,
\end{equation}
For simplicity we then choose $b_0^2 Q^2 =4\nu$, or equivalently $\bbar_0^2
Q^2= e^{\gae} \nubar$. 

The inverse Mellin transform with respect to $\nu$ is similar to the
that for $\tau_Q$, and we obtain as our final answer (remembering that 
$\nu$ was conjugate to $\tau/2$)
\begin{align}
\Sigma_N(\tau_E ) &\simeq 
  e_q^2 q_N(\tau_E  Q^2) \frac{\Gamma(1 - R'_U)}{\Gamma(1 +
  R' + R'_U)} \exp\left[-
  R\left(\frac{2e^{\gae}}{\tau_E } \right) -
  R_U\left(\frac{2e^{2\gae}}{\tau_E } \right) 
            \right]\\
            \label{eq:taueFinal}
            &\simeq e_q^2 q_N(\tau_E  Q^2) \frac{\Gamma(1 - R'_U)}{\Gamma(1 +
              R' + R'_U)}  \exp\left[-
  R\left(\frac{1}{\tau_E } \right) -
  R_U\left(\frac{1}{\tau_E } \right)  \right. \\ 
    &\qquad\qquad\qquad\qquad\qquad\qquad\qquad\qquad\qquad
  -  (\gae + \ln 2)R' - (2\gae + \ln2)R'_U \nonumber
            \bigg].
\end{align}
As before, $R'=R'(1/\tau)$.

\section{The final result}
\label{sec:finalres}

It is possible to go one step further in the resummation, namely the
determination of constant terms at $\cO{\as}$. These are terms
independent of $\tau$ which essentially multiply the resummed answer.
Their function is to compensate for mismatches between the independent
gluon emission pattern (as applied to a single gluon) and the actual
emission pattern, usually in \emph{corners} of phase space, regions
not enhanced by any logs. The corners that can arise are the region of
hard, non-collinear emission, and, for the thrust, the region of
collinear, non-soft emissions with $k_t^2 \sim \tau Q^2$. There are
also other origins for such terms, such as certain approximations made
in the expression for the thrust, and contributions from virtual
factorisation scheme corrections.

The contribution to $F_2$ from events with $1-T<\tau$ is therefore
given by the following expression:
\begin{multline}\label{eq:FinalForm}
  \Sigma(x,Q^2,\tau) = x\Bigg[\sum_{q,\qbar} e_q^2 \left(q(x,\tau Q^2) +
  \frac{\as(Q^2)}{2\pi} C_{1q} \otimes q(x,Q^2)\right) \\
 + \left(\sum_{q,\qbar} e_q^2\right)  \frac{\as(Q^2)}{2\pi} C_{1g} \otimes
 g(x,Q^2) \Bigg]  
e^{L g_1(\as \beta_0 L) + g_2(\as \beta_0 L)},
\end{multline}
where $L=\ln 1/\tau$ and the convolutions are defined as
\begin{equation}
  \label{eq:conv}
  C_{1q} \otimes q(x,Q^2) = \int_x^1 \frac{dz}{z}\, C_{1q}(z) \,
  q\left(\frac{x}{z},Q^2\right),
\end{equation}
and we have written the non-$x$ dependent leading and subleading
logarithms as $Lg_1(\as \beta_0 L)$ and $g_2(\as \beta_0 L)$
respectively, whose forms follow from eqns.~\eqref{eq:tauqFinal} and
\eqref{eq:taueFinal} and the expressions for $R$ and $R_U$ in
appendix~\ref{app:twoloopR}.  Specifically, for $\tau_Q$ we have
(where it should be noted that the $R_1$ and $R_2$ are defined with
different arguments from $R$)
\begin{subequations}  \label{eq:tauQg12}
\begin{align}
  g_1(\as \beta_0 L)  &= -2R_1(\as \beta_0 L),\\
  g_2(\as \beta_0 L) &= -2R_2(\as \beta_0 L)- 2\gae R' - \ln
  \Gamma(1+2R'),
\end{align}
\end{subequations}
and for $\tau_E$,
\begin{subequations}
  \label{eq:tauEg12}
\begin{align}
  g_1(\as \beta_0 L) &= -R_1(\as \beta_0 L) - R_{U1}(\as \beta_0 L),\\
  g_2(\as \beta_0 L) &= -R_2(\as \beta_0 L) - R_{U2}(\as \beta_0 L) \\ 
  & \nonumber \qquad\qquad\qquad\qquad - (\gae + \ln 2)(R'+R_U') -
  \gae R_U' + \ln \frac{\Gamma(1 - R'_U)}{\Gamma(1 + R' + R'_U)}\,.
\end{align}
\end{subequations}
It is useful to have their order by order expansions,
\begin{subequations}
\begin{align}
  L g_1(\as \beta_0 L) &= \sum_n G_{n,n+1} \left(\frac{\as}{2
      \pi}\right)^n L^{n+1}, \\ 
  g_2(\as \beta_0 L) &= \sum_n G_{n,n} \left(\frac{\as}{2
      \pi}\right)^n L^{n},
\end{align}
\end{subequations}
where the $\cO{\as}$ and $\cO{\as^2}$ coefficients are given in
table~\ref{tab:Gs}.

\TABLE{
\begin{tabular}{|c|c|c|c|c|}\hline
         & $G_{12}$ & $G_{11}$ & $G_{23}$ & $G_{22}$  \\ \hline
$\tau_Q$ & $-2\cf$  & $3\cf$   & $2\pi\beta_0 G_{12}$ &
         $ -\frac43\pi^2\cf^2 + \left(\frac{\pi^2}{3} -
           \frac{169}{36}\right)\ca\cf  + \frac{11}{18}\cf\nf$ 
         \\ \hline 
$\tau_E$ & $-\frac32\cf$ & $(3-3\ln2)\cf$& $\frac{16\pi}{9}\beta_0 G_{12}$&
         \begin{minipage}{0.45\textwidth}
           $-\frac23\pi^2 \cf^2 + 
           (\frac{\pi^2}{4} - \frac{17}{6} - \frac{22}{3}\ln2)\ca\cf $
           \\
           \hfill  $+(\frac13 + \frac43\ln2)\nf\cf$
         \end{minipage}
     \\ \hline
\end{tabular}
\label{tab:Gs}
\caption{Coefficients in the resummation.}
}

The expressions for the $C_1$'s are given, in the DIS and $\MSbar$
factorisation schemes, in appendix~\ref{app:calcLO}. These constant
terms have been associated with a scale $Q^2$, both in $\as$ and in
the parton distributions. Strictly such an identification of the scale
is not unique: indeed $C_1$ can come from pieces with a variety of
scales, however the error made in neglecting this fact amounts to
terms $\as^{n+1}L^n$ and thus is beyond our accuracy. A further point
worth mentioning relates to the cutoff on the visible energy, $\elim$.
For both $\tau_Q$ and $\tau_E$, $C_1$ is independent of $\elim$, as
long as events with less than $E<\elim$ in the current hemisphere are
given a thrust of 0: this is because a modification of $\elim$ just
redistributes thrust configurations between some finite value of
$\tau$ and $\tau=1$; but the small $\tau$ region is not affected. (For
certain other variables, such as the thrust measured with respect to
the thrust axis and normalised to $E$, this would not be the case).

\subsection{Comparison with NLO computations}
\label{sec:NLO}

A valuable test of the resummation is to expand our answers to second
order in $\as$ and compare the result with fixed order predictions
from programs such as DISASTER++ \cite{DR} and DISENT \cite{DT}.
Though agreement does not guarantee the correctness of the
resummation, it is nevertheless a non-trivial check.

For the comparison we simply look at the second order contribution to 
\begin{equation}
  \frac{1}{{\sigma}_{0,N}(Q^2)} \frac{d\Sigma_N (Q^2,\tau)}{d \ln
    \frac{1}{\tau}}\,,
\end{equation}
where ${\tilde\sigma}_0$ is just the $\cO{\as^0}$ part of the
structure function moment $F_{2,N}(Q^2)$.  Moment space with respect
to $x$ is used to reduce the various convolutions (for the $C_1$'s and
the change of scale of the parton distribution) to products. At order
$\as^2$, the terms that we control go as $\as^2 L^3$, $\as^2L^2$ and
$\as^2L$ and their coefficients can be determined by expanding
\eqref{eq:FinalForm}.  Therefore the difference between our expanded
resummed answer and the exact second order coefficient is a term which
for large $L$ is at most a constant.  Figure~\ref{fig:tauQplain} shows
the expanded resummed and exact second order results (from the fixed
order Monte Carlo program DISASTER++) in the quark and gluon channels.

The shape of the resummed and exact results are clearly similar and
the difference between them does seem to tend to a constant at small
$\tau$. This is visible more clearly in figure~\ref{fig:tauQdiff}
which shows the difference between the two results. Also shown is the
difference between the current version (0.1) of DISENT and the
resummed answer. In the gluon channel the difference is clearly
inconsistent with a constant: it seems to go as $\as^2L^2$. In the
quark channel there is also evidence of similar problems, with a
significance of a few (between 3 and 5) standard deviations.  Results
are shown for $\tau_E$ in figure~\ref{fig:tauEdiff}, where the same
pattern of successes and problems is seen.

The origin of the discrepancy in DISENT has yet to be elucidated.
However it has been possible to identify that the disagreement lies in
pieces proportional to $\cf^2$ (quark channel) and $\ca \tr$ (gluon
channel). Furthermore the fact that, in the gluon channel, the
disagreement is at the level of a $\as^2L^2$ term in the distribution
($\as^2L^3$ in the integrated distribution) suggests that the problem
lies with a term involving one soft and two collinear divergences. In
the gluon channel there should be \emph{no} term $\as^2L^2$ with
colour factor $\ca\tr$ (due to angular ordering, soft gluon radiation
in the region relevant to the thrust is associated with the colour
factor $\cf$ because it is radiated at an angle larger than that of
the quark emitted in the photon-gluon fusion).

Above we have said that the difference between the resummed and
DISASTER++ is consistent with a constant. This should be qualified: in
the gluon channel, the largest-allowed non-constant is quite small,
about $\pm 0.2 (\as/2\pi)^2 L$ for both $\tau_Q$ and $\tau_E$. In the
quark channel the statistical error on the results is much larger,
because the largest term goes as $\as^2L^3$ (in the gluon channel it
goes as $\as^2L^2$), and for a given number of events the statistical
error is proportional roughly to the largest term. The largest allowed
non-constant term here would go as $\sim \pm 3 (\as/2\pi)^2 L$.  In
both the quark and gluon channels the uncertainty represents about
$10\%$ of the coefficient of the $(\as/2\pi)^2 L$ term from the
resummation.

The DISASTER++ results have been obtained\footnote{Using the Condor
  system \cite{condor} to distribute the processing across INFN
  computers spread across the whole of Italy} with the equivalent of
50 days' time on a machine with a SPEC CFP95 \cite{specfp} of about 20
(or roughly equivalently, 500 MIPS). Thus significantly higher
precision is not really feasible.  (DISENT, for comparable errors, is
an order of magnitude faster, so when the the problems with DISENT
have been resolved a more stringent comparison will be feasible;
DISENT also shows less dependence on the internal cutoff, required in
order to avoid floating point errors).

\FIGURE{
  \epsfig{file=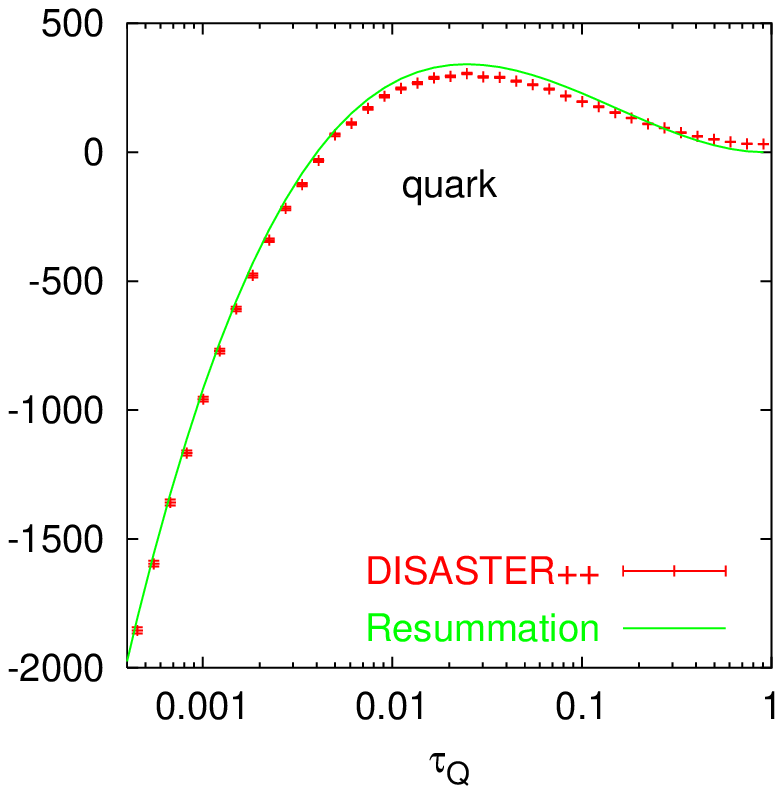,width=0.47\textwidth}
  \epsfig{file=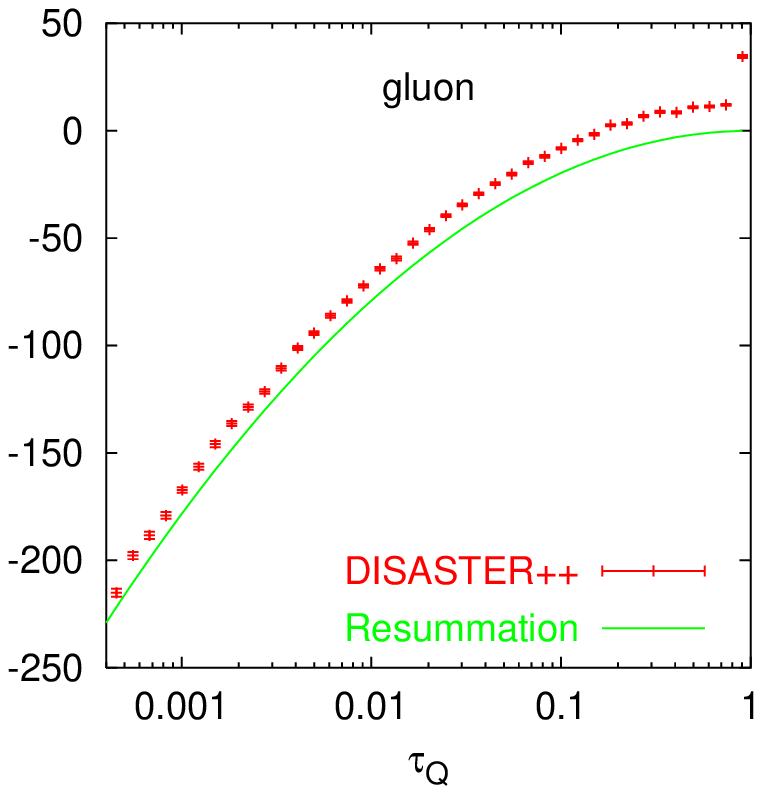,width=0.47\textwidth}
  \caption{For $\tau_Q$, a comparison of the coefficient of
    $(\as/2\pi)^2$ from the expanded resummed answer and that from
    DISASTER++. Shown for $N=1.4$.}
  \label{fig:tauQplain}
  }

\FIGURE{
  \epsfig{file=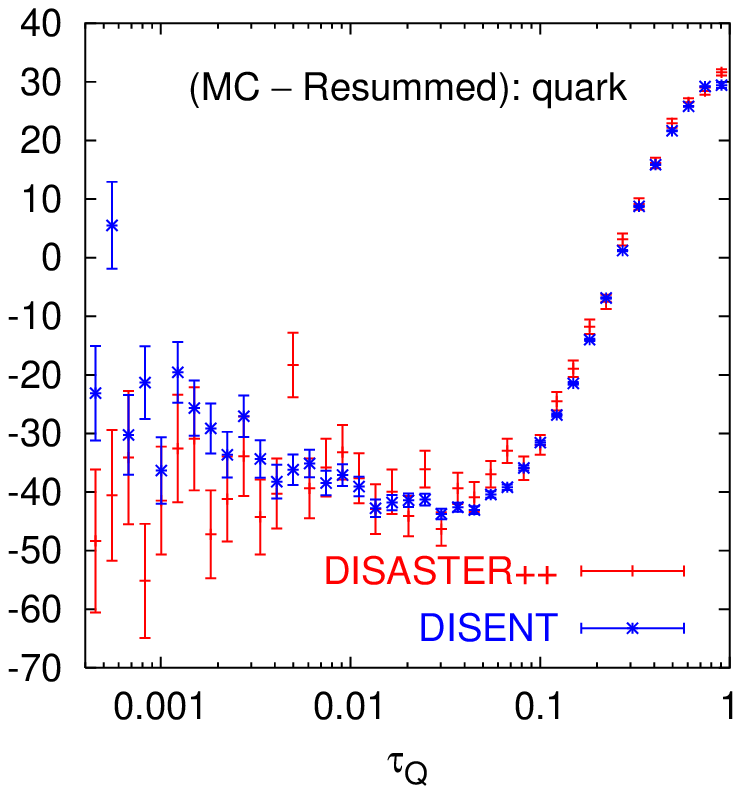,width=0.47\textwidth}
  \epsfig{file=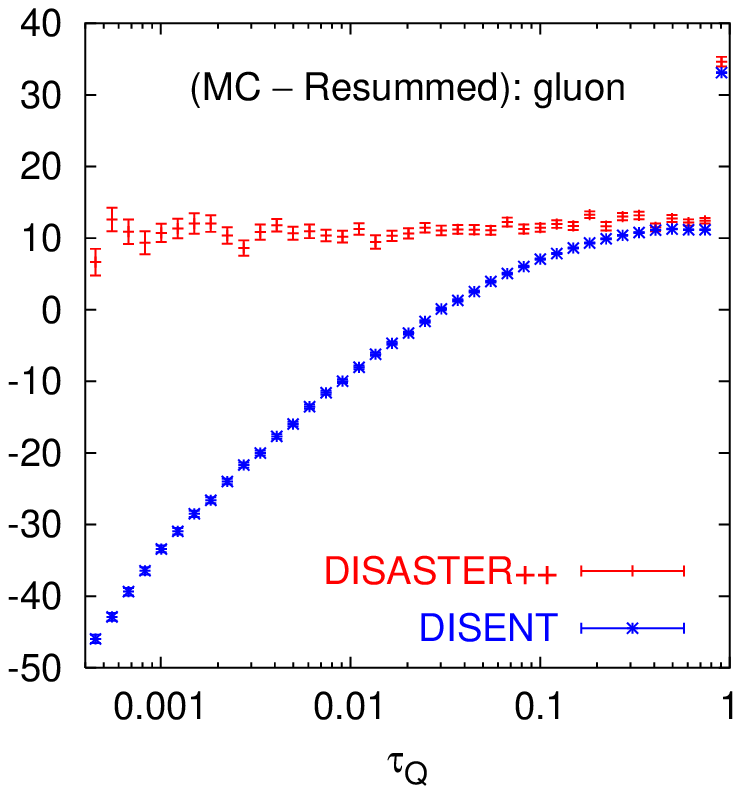,width=0.47\textwidth}
  \caption{For $\tau_Q$, the difference between the  second order
    coefficient from the Monte Carlo programs (DISASTER and DISENT)
    and that from the expanded resummed answer. Shown for $N=1.4$.} 
    \label{fig:tauQdiff}
  }

\FIGURE{
  \epsfig{file=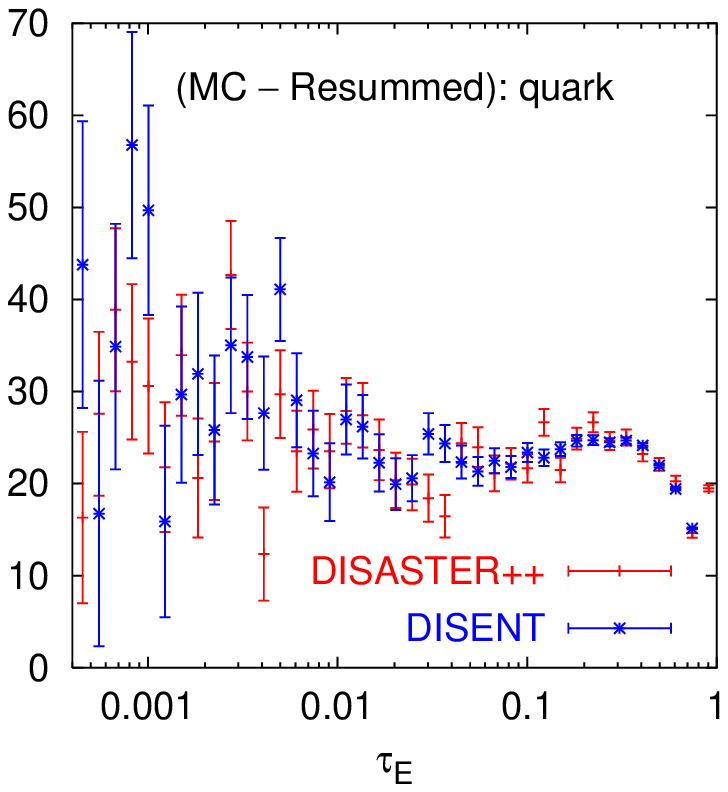,width=0.47\textwidth}
  \epsfig{file=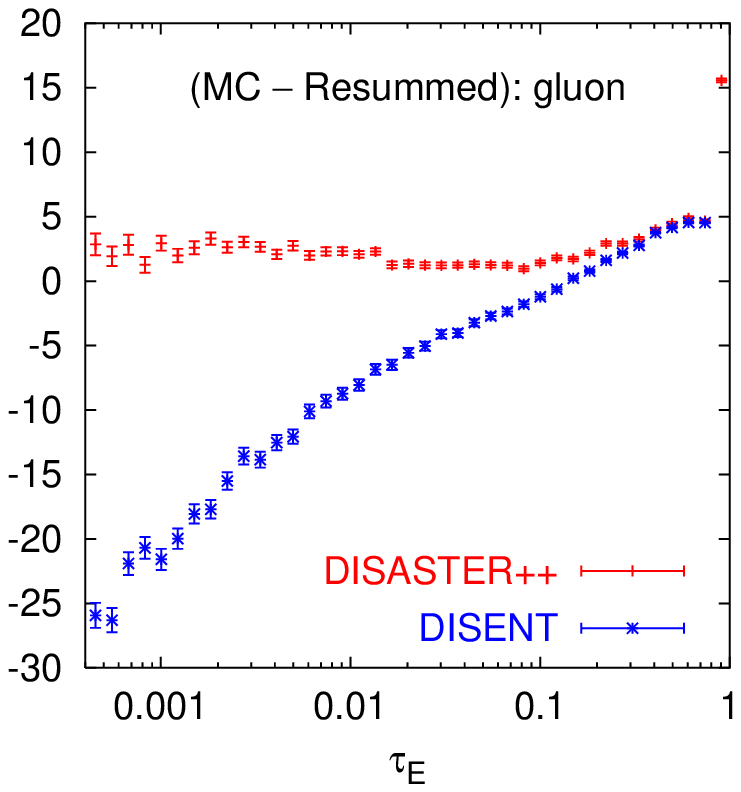,width=0.47\textwidth}
 \caption{For $\tau_E$, the difference between the  second order
   coefficient from the Monte Carlo programs and that from the
   expanded resummed answer. Shown for $N=1.7$.}
 \label{fig:tauEdiff}
 }

\subsection{Matching}
\label{sec:match}

The above form for the answer is satisfactory for very small $\tau$.
However in practice one wishes to fit to data over a range of $\tau$.
In $\ee$ sophisticated methods have been developed for combining the
$\cO{\as^2}$ and the resummed calculations in order to have an
expression which can be applied over the whole range of data
\cite{CTTW}. This procedure is usually called matching. It essentially
involves two elements: the removal of terms which would otherwise lead
to double counting (\ie logarithms present in both the fixed-order and
resummed answers) and the exponentiation of the subleading logarithms
($\as^2L$ which is associated with the coefficient $G_{21}$ of
$g_3(\as \beta_0 L)$) which are present in the
fixed-order calculation.

In DIS the situation is a little more delicate because the term
$\as^2L$ which would normally be associated with $G_{21}$ is difficult
to exponentiate --- for example some of it might arise because the
parton distributions convoluted with the $C_1$'s should really have
been evaluated at a different scale, or because of contributions from
the NLO splitting functions. Some of these contributions exponentiate
only in moment space (with respect to $x$) and as a matrix in flavour
space. Thus the simple exponentiation carried out in $\ee$ is not
appropriate here.

However such an exponentiation is not strictly necessary. What is
necessary \cite{CTTW} is to ensure that terms such as $\as^2L$ are not
independent of the resummation, but rather a subleading modification
of the resummation; thus our final answer must contain not $\as^2L$ on
its own, but $\as^2L \exp(Lg_1(\as \beta_0 L) + \ldots)$.

As yet we have not yet carried out a detailed study of possible
matching schemes. However for completeness we do include a preliminary
proposal for matching, which satisfies the requirements discussed
above and which will be examined in more depth elsewhere
\cite{ADSfuture}. 

Since these matched expressions will be directly compared to
data, it is useful at this point to switch to considering the
cross section, differential in $x$ and $Q^2$, for $1-T$
to be less than $\tau$. For the resummed differential cross section,
which we will represent by $\dsig_R$, we have
\begin{equation}
  \label{eq:xsect}
  \dsig_R \equiv
  \frac{d^2 \sigma_R(x,Q^2,\tau)}{dx dQ^2} = \frac{4\pi\alpha^2}{xQ^4} 
  \frac{(1 + (1-y)^2)}{2}\Sigma(x,Q^2,\tau)\,.
\end{equation}
For brevity it will turn out to be useful to define
the following matrices:
\begin{equation}
  \label{eq:matrices}
  \bq(x) = \left(
    \begin{array}{c}
      q_u(x)\\
      q_{\bar u}(x) \\
      \vdots\\
      g(x)
    \end{array}
  \right),\qquad
  \bP(x) = \left(
    \begin{array}{cccc}
      P_{qq}(x) & 0 & \cdots & P_{qg}(x)\\
         0      & P_{qq}(x) & & \\
      \vdots & & \ddots & \\
      P_{gq}(x) & & & P_{gg}(x)
    \end{array}
  \right),
\end{equation}
and
\begin{align}
  \bC_0^T(x) &= \frac{4\pi\alpha^2}{Q^4} 
  \frac{(1 + (1-y)^2)}{2} \left(
    \begin{array}{c}
      e_u^2 \delta(1-x) \\ e_u^2 \delta(1-x) \\ \vdots \\ 0
    \end{array}
  \right),\\
  \bC_1^T(x) &= \frac{4\pi\alpha^2}{Q^4} 
  \frac{(1 + (1-y)^2)}{2} \left(
    \begin{array}{c}
      e_u^2 C_{1,q}(x) \\ e_u^2 C_{1,q}(x) \\ \vdots \\ \sum_{q,\qbar} e_q^2
      C_{1,g}(x) 
    \end{array}
  \right).
\end{align}
With this notation, and using eqs.~\eqref{eq:FinalForm} and
\eqref{eq:xsect}, the resummed differential cross section becomes
\begin{equation}
  \label{eq:FinalFormMat}
  \dsig_R(x,Q^2,\tau) = \left[ \bC_0\otimes \bq(x,\tau Q^2) + \asb(Q^2)
    \bC_1\otimes \bq(x,Q^2) \right] 
  e^{L g_1(\as \beta_0 L) + g_2(\as \beta_0 L)},
\end{equation}
where we have introduced $\asb = \as/2\pi$. 

The form that we propose for the matching is the following
\begin{multline}
\label{eq:matched}
\dsig(x,Q^2,\tau) = \dsig_{R} + \Big[ 
\asb \left(\dsig_E^{(1)} - \dsig_{R}^{(1)}\right)
            \\   + \asb^2\left(\dsig_E^{(2)} -
               \dsig_{R}^{(2)} - (\dsig_E^{(1)} -
               \dsig_{R}^{(1)})(L^2 G_{12} + L G_{11})
             \right)\Big]
             e^{ Lg_1(\as \beta_0 L) + g_2(\as \beta_0 L)},
\end{multline}        
where $\dsig_E^{(n)}$ is the coefficient of $\asb^n$ in the exact
result (as determined from Monte Carlo programs) and $\dsig_R^{(n)}$
is the coefficient of $\asb^n$ in the resummed answer:
\begin{align}
  \dsig_R^{(1)} &= \left[\bC_0\left(G_{12} L^2 + G_{11}L\right) 
     -\bC_0\otimes \bP L +  \bC_1\right] \otimes \bq(x,Q^2)\,, \\
  \dsig_R^{(2)} &= 
  [
    \bC_0 ( 
      1/2\,G_{12}^2 L^4
      +\left( G_{23}+G_{11}G_{12} \right){L}^{3}
      + ( 1/2\,G_{11}^2        +G_{22}){L}^{2} )
         +\bC_0\otimes (-G_{12}\bP L^3  
        \nonumber\\&\qquad  \label{eq:expsn2}
        + (\half\,\bP\otimes\bP
         -G_{11}\bP
        -\,\pi\beta_0 \bP )L^2)
      +\bC_1 (G_{12} L^2 + G_{11}L) 
  ] \otimes \bq(x,Q^2)\,.
\end{align}
For these expressions to be correct, in eq.~\eqref{eq:FinalFormMat}
the evolution of the parton distributions to scale $\tau Q^2$ must be
carried out from scale $Q^2$ with just the leading log DGLAP equations
(in any case this is all that can be guaranteed by the resummation
procedure). If one wants to use the NLL DGLAP equations, one may, but
then eq.~\eqref{eq:expsn2} should be modified accordingly.

Finally we point out that the effect of power corrections on the
distribution should just be to \emph{shift} the distribution to larger
$\tau$ by an amount equal to the shift of the mean values as
calculated\footnote{With the proviso that the Milan factor
  \cite{DLMS2loop,DLMSmilan} should take on the updated value of
  $1.49$ for three light flavours \cite{DasMagSmy,YuriProc}.} in
\cite{DWDIS,DWmilan}, as is the case in $\ee$
\cite{KorSter,DWtdist,DLMSmilan}.

\section{Conclusions}
\label{sec:conclusions}

We have determined the resummed distributions for two definitions of
thrust, in the region $T\to1$. For $\tau_Q$ the answer is almost
identical to the $\ee$ result, except that the answer is multiplied by
a parton distribution, which is evaluated not at scale $Q^2$ but at
scale $\tau Q^2$. This arises because placing a limit on the thrust
amounts, in the remnant collinear region, to placing a limit on the
emitted transverse momentum.  The thrust normalised to the energy in
$\hc$ also has this dependence on the quark distribution at $\tau
Q^2$. But it differs from $\tau_Q$ in that the coefficient of the
leading double logarithm, $G_{12}$ is only $3/4$ of that of $\tau_Q$,
the reason being that the two normalisations depend differently on
emissions in the remnant hemisphere.

These resummed expressions have been expanded and compared to the
predictions from the fixed-order Monte Carlo programs DISASTER++ and
DISENT. We find reasonable agreement with DISASTER++, but a definite
discrepancy compared to DISENT. The discrepancy is most visible in the
gluon channel, but there is evidence that there might be a problem
also in the quark channel.

We have also given a `matching' prescription for combining the
resummed and fixed order predictions: it satisfies the basic
properties of being correct both to second fixed order and for the
leading and sub-leading logarithms, and additionally of not having
`large' (constant or logarithmic) pieces left over at small $\tau$. 

There remain a certain number of other DIS event-shape variables that
can be resummed. They include the broadening, the $C$-parameter and
the thrust measured with respect to the thrust axis and the jet mass.
These will be examined elsewhere \cite{ADSfuture}.

\acknowledgments

We benefited much from continuous discussions of this and related
subjects with Stefano Catani, Yuri Dokshitzer, Pino Marchesini, Mike
Seymour and Bryan Webber. We would also like to thank Francesco Prelz
for Condor support. One of us (GPS) would like to thank the Milan
group for hospitality and the use of computer facilities during the
writing of this article.

\appendix

\section{Leading order calculations}
\label{app:calcLO}

Here we give the explicit expressions for the functions $C_{1q}(z)$
and $C_{1g}(z)$ entering the result, eq.~\eqref{eq:FinalForm}, for the
integrated thrust distribution.  Essentially, they are obtained by
carrying out the full leading-order calculation, taking the limit of
small $\tau$, and then subtracting the pieces with a logarithmic
dependence on $\tau$.

We give results for the $F_2$ part only, in accord with
eq.~\eqref{eq:FinalForm}, since the component of the cross section
contributing to $F_L$ is suppressed by a power of $\tau$ at small
$\tau$. In the DIS factorisation scheme, the results for $\tau_Q$ are
as follows:
\begin{align}
C_{1q}^{\DIS} (z) &= -C_F \left[\frac{1+z^2}{1-z} \ln
  \frac{1}{z} + 
\frac{1+2 z -6 z^2}{2 \, (1-z)_+} \right]
,\\
C_{1g}^{\DIS} (z) &= -T_R
\left[\left(z^2 + (1-z)^2\right) \ln \frac{1}{z}
- \left(1 -6 z \left(1 -z \right)\right) \right]
\, .
\end{align}
The plus prescription has its usual meaning. The equivalent results in
moment space (with respect to $z$) are
\begin{align}
  C_{1q,N}^{\DIS} 
  &=  -C_F \left[\frac{5}{N+1} -\frac{1}{N^2
      \left(N+1\right)^2}
    +\frac{3}{2} \left(\psi(N)+\gamma\right) +2 \psi'(N)\right]\, ,\\
  C_{1g,N}^{\DIS} &= -T_R \left[\frac{2}{(N+2)^2} -
    \frac{2}{(N+1)^2} + \frac{1}{N^2} - \frac{1}{N} + \frac{6}{N+1}
    -\frac{6}{N+2}\right],
\end{align}
where $\psi(N) = d/dN \ln \Gamma(N)$. 

The analogous results for $\tau_E$ are somewhat more complicated,
because of the more involved nature of the phase space integrations
for that variable:
\begin{align}
C_{1q}^{\DIS} (z) =&  -C_F \left[ \frac{1+z^2}{1-z} 
\ln \frac{1}{z} 
+ (1+z^2) \left(\frac{\ln (2(1-z))}{1-z}\right)_+ 
+\frac{ 1 + 4\ln2 +2 z -6 z^2}{2 \left(1-z\right)_+} \right. 
\nonumber\\ 
 & \left.   
\qquad\qquad\qquad-(1 + z) \ln 2 
 +\frac{3}{2} \left(\ln^2 2 -\ln 2 \right) \delta(1-z)\right]
, \\
C_{1g}^{\DIS} (z) =& 
-T_R \left[\left(z^2+\left(1-z\right)^2\right) 
\ln \frac{2(1-z)}{z}
+ \left(-1 + 6 z -6 z^2 \right)\right] .  
\end{align}
The results in moment space are 
\begin{align}
C_{1q,N}^{\DIS}
= & -C_F \left[\frac{\pi^2}{6}+\frac{2}{N}+\frac{3}{N+1}
  +  \frac{3}{2} \psi(N) +\frac{3}{2}\gamma +\gamma \psi(N)+\gamma 
\psi(N+2)+\gamma^2+\right.\nonumber\\
 & + \left. \frac{1}{2} (\psi'(N)+\psi'(N+2))+
\frac{1}{2} ((\psi(N))^2 +(\psi(N+2)^2)) + \right. \nonumber\\ 
 & + \left. \frac{3}{2} (\ln^2 2-\ln 2)-\ln 2 \left(2 
\left(\psi(N)+\gamma \right) +\frac{1}{N}+\frac{1}{N+1} \right)\right],
\\
C_{1g,N}^{\DIS}
=&-T_R \left[-\frac{1}{N} \left(1+\psi(N)+\gamma\right)+
\frac{2}{N+1} \left(3+\psi(N+1)+\gamma\right)-\right.\nonumber\\ 
& - \left.  \frac{2}{N+2}\left(3+\psi(N+2)+\gamma\right)  + 
   \ln 2 \left(\frac{1}{N}-\frac{2}{N+1}+ \frac{2}{N+2}
   \right)\right]. 
\end{align}

Additionally to go from the DIS to the $\MSbar$ factorisation scheme
result, it suffices to add the $\cO{\as}$ $\MSbar$-scheme $F_2$
coefficient function to the above results:
\begin{align}
  C_{1q}^{\MSbar} =&   C_{1q}^{\DIS} + 
  \cf\left[
    2 \left(\frac{\ln(1-z)}{1-z}\right)_+
     - \frac32\frac1{(1-z)_+} - (1+z)\ln(1-z) \right.\nonumber\\
     &\left. \qquad\qquad-\frac{1+z^2}{1-z}\ln z + 3 + 2z 
     - \left(\frac{\pi^2}{3} + \frac92\right)\delta(1-z)
  \right],\\
  C_{1g}^{\MSbar} =&   C_{1q}^{\DIS} + 
  T_R\left[
    \left((1-z)^2+z^2\right)\ln \frac{1-z}{z} - 8z^2 + 8z - 1
  \right].
\end{align}

\section{Inverse Mellin transforms}
\label{app:invmellin}

In \cite{DMS} the following operator technique was developed to help
carry out the Mellin transforms. We write 
\begin{equation}
  e^{-\cR(\nu)} = \left. e^{ - \cR(e^{-\da})}\, \nu^{-a}
  \right|_{a=0} 
\end{equation}
The inverse Mellin transform is given by 
\begin{equation}
\label{eq:Sigop}
 \Sigma(\tau) =  \int\frac{d\nu}{2i\pi\nu}\>e^{\tau\nu}\> 
 e^{ - \cR(e^{-\da})}\, \nu^{-a}
 = \left. e^{-\cR(e^{-\partial_a})}\>
\frac{1}{\Gamma(1+a)}\>\left(\frac{1}{\tau}\right)^{-a} 
\right|_{a=0}.
\end{equation}
Using the identity
\begin{equation}
   \left. e^{-\cR(e^{-\partial_a})}\> x^{-a}g(a) \right|_{a=0} \>=\> 
   \left.  e^{-\cR(xe^{-\partial_a})} \>g(a)\right|_{a=0}
   \>,\label{eq:ident}
\end{equation}
we can absorb a power of $(1/\tau)$ into the argument of
the radiator
\begin{equation}
  \Sigma(\tau) =
  \left. e^{-\cR\left(\frac{1}{\tau}e^{-\partial_a}\right)}\>  
\frac{1}{\Gamma(1+a)}\>  \right|_{a=0}.
\end{equation}
Performing the logarithmic expansion of the radiator:
\begin{equation}
 -\cR\left(xe^{-\partial_a}\right) \>=\> -\cR(x) + \cR'(x)\partial_a
  -\half\cR''(x)\partial_a^2+ \ldots \>, 
\end{equation}
we obtain that the action of the operator on a function which is {\em
  regular}\/ in the origin reduces to substituting $\cR'(x)$ for $a$,
while $\cR''(x)=\cO{\as}$ and higher derivatives produce negligible
corrections:
\begin{equation}
  \Sigma(\tau) = e^{-\cR\left(\frac{1}{\tau}\right)}\>  
\frac{1}{\Gamma(1+\cR')}\, \qquad \cR' =
\cR'\left(\frac{1}{\tau}\right) \,.
\end{equation}

\section{Two-loop radiator integrals}
\label{app:twoloopR}

There are two integrals which have to be evaluated:
\begin{equation}
  \label{eq:intupper}
  R_U(\nu) = \int_{Q^2\nu^{-1}}^1 \,\frac{d k_t^2}{k_t^2}
  \frac{\as(k_t^2) \cf}{2\pi} \left( \ln
  \frac{Q^2}{\kt^2}  - \frac{3}{2}\right),
\end{equation}
and
\begin{equation}
  R_L(\nu) = \label{eq:intlower}
  \int_{Q^2\nu^{-2}}^{Q^2\nu^{-1}}\, \frac{d k_t^2}{k_t^2} 
  \frac{\as(k_t^2) \cf}{2\pi}
  \ln (\nu^2 \kt^2/Q^2)\,.
\end{equation}
For the results to be correct at two-loop accuracy it is necessary
that $\as$ be the CMW or Physical scheme coupling \cite{CMW} and that
the running of $\as$ be taken into account to two-loop level. The CMW
scheme coupling is related to the $\MSbar$ coupling through
\begin{equation}
  \as^{\mathrm{CMW}} =  \as^{\MSbar}  \left(1 + \frac{\as}{2\pi}
    K\right), \qquad K = \ca\left(\frac{67}{18} -
    \frac{\pi^2}{6}\right)
  - \frac{5}{9}\nf\,,
\end{equation}
and the two-loop running of the coupling is reproduced to sufficient
accuracy by the expression
\begin{equation}
  \label{eq:alpha2loop}
  \as(\mu^2) = \as(Q^2) \left[\frac1{1- \lambda} 
    - \frac{\beta_1}{\beta_0} \frac{\as(Q^2) \ln
      (1-\lambda)}{(1-\lambda)^2}
    \right], \qquad\quad \lambda = \as(Q^2) \beta_0 \ln
    \frac{Q^2}{\mu^2}\,,
\end{equation}
where 
\begin{equation}
  \label{eq:betas}
  \beta_0 = \frac{11\ca - 2\nf}{12\pi},\qquad\quad 
  \beta_1 = \frac{17\ca^2 -5\ca\nf - 3\cf\nf}{24\pi^2}\,.
\end{equation}
We split the results into leading and subleading logarithms:
\begin{equation}
  \label{eq:split}
  R_U(\nu) = \ln \nu \cdot R_{U1}\left(\as(Q^2) \beta_0 \ln \nu\right) +
  R_{U2}(\as \beta_0 \ln \nu) 
\end{equation}
and analogously for $R_L$ and $R=R_U+R_L$. The results are as follows:
\begin{subequations}\label{eq:RU}
\begin{align}
  R_{U1}(\lambda) =& \frac{\cf}{2\pi\beta_0\lambda} \left
    [-\lambda-\ln{(1-\lambda)}\right]\,,\\
  R_{U2}({\lambda})=&  \frac{3\cf}{4\pi \beta_0} \ln(1-\lambda)
  +\frac{K\cf [\lambda + (1-\lambda)\ln(1-\lambda)]}
  {4\pi^2\beta_0^2(1-\lambda)} 
  \\\nonumber &\qquad\qquad\qquad +
  \frac{\cf\beta_1}{2\pi\beta_0^3} \left[ -\frac{\lambda + \ln
      (1-\lambda)}{1-\lambda} - \frac12 \ln^2{(1-\lambda)} \right] ,
\end{align}
\end{subequations}
and
\begin{subequations}
\begin{align}\label{eq:RL}
  R_{L1}(\lambda) =& \frac{\cf}{2\pi\beta_0\lambda}\left[\lambda+
    (1-2\lambda) \ln{\left(\frac{1-2\lambda}{1-\lambda}\right)}
  \right], 
  \\
  R_{L2}(\lambda) =& \frac{K\cf \left[
      (1-\lambda)\ln\frac{1-\lambda}{1-2\lambda} - \lambda \right]}
  {4\pi^2\beta_0^2(1-\lambda)} 
  + \frac{\cf \beta_1}{2 \pi
    \beta_0^3}\left[ \frac{\lambda + (2\lambda-1) \ln
      (1-\lambda)}{1-\lambda} \right.\\\nonumber 
  & \qquad\qquad\qquad\qquad\left.+ \frac12 \ln^2(1-2\lambda)- \frac12
    \ln^2(1-\lambda) + \ln(1-2\lambda) \right],
\end{align}
\end{subequations}
and for the sum $R = R_U + R_L$,
\begin{subequations}\label{eq:RUL}
\begin{align}
  R_1(\lambda) =& \frac{\cf}{2\pi\beta_0\lambda}
  \big[(1-2\lambda)\ln{(1-2\lambda)} - 2(1-\lambda) \ln{(1-\lambda)}
  \big],\\
  R_2(\lambda) =& \frac{3\cf}{4\pi \beta_0} \ln(1-\lambda) +\frac{K
    \cf [2\ln(1-\lambda) - \ln(1-2\lambda)]}{4\pi^2\beta_0^2}
  \\\nonumber &+\frac{\cf \beta_1}{2 \pi
    \beta_0^2}\left[\ln(1-2\lambda)-2\ln(1-\lambda)
    +\frac{1}{2}\ln^2(1-2\lambda)-\ln^2(1-\lambda)\right ] .
\end{align}
\end{subequations}
We also need the expressions for the derivatives of the $R$'s
with respect to $\ln \nu$:
\begin{subequations}
  \label{eq:Rprime}
  \begin{align}
  R'_U &= \frac{\cf}{2\pi \beta_0} \frac{\lambda}{1-\lambda}, \\
  R'_L &= \frac{\cf}{2\pi \beta_0}\left(
    2\ln \frac{1-\lambda}{1-2\lambda} 
    -\frac{\lambda}{1-\lambda} \right),\\
  R' &= \frac{\cf}{\pi \beta_0} \ln
  \frac{1-\lambda}{1-2\lambda} ,
  \end{align}
\end{subequations}
where $\lambda = \alpha_s \beta_0 \ln \nu$. A final point relates to
scale changes. If in $\ln \nu R_1(\lambda)$ we change the scale of
$\as$ from $Q^2$ to $\mu^2$, then $R_2$ gets modified as follows:
\begin{equation}
  \label{eq:scalechange}
  R_2 \to R_2 + \lambda \ln \frac{\mu^2}{Q^2} \left(R' - R_1\right)\,,
\end{equation}
and analogously for $R_U$ and $R_L$.


\end{document}